\documentclass[reprint,aps,prd,amsmath,amssymb,floatfix,nofootinbib,preprintnumbers]{revtex4-2}

\usepackage{graphicx}
\usepackage[dvipsnames]{xcolor}
\usepackage{hyperref}
\hypersetup{colorlinks, urlcolor=BlueViolet, citecolor=Plum, linkcolor=PineGreen}

\usepackage{bm}
\usepackage{cleveref}

\begin{document}

\title{Exploring light dark matter with the Migdal effect in hydrogen-doped liquid xenon}

\author{Nicole F.~Bell}
\email{n.bell@unimelb.edu.au}
\affiliation{ARC Centre of Excellence for Dark Matter Particle Physics, School of Physics, The University of Melbourne, Victoria 3010, Australia}

\author{Peter Cox}
\email{peter.cox@unimelb.edu.au}
\affiliation{ARC Centre of Excellence for Dark Matter Particle Physics, School of Physics, The University of Melbourne, Victoria 3010, Australia}

\author{Matthew J.~Dolan}
\email{dolan@unimelb.edu.au}
\affiliation{ARC Centre of Excellence for Dark Matter Particle Physics, School of Physics, The University of Melbourne, Victoria 3010, Australia}

\author{Jayden L.~Newstead}
\email{jnewstead@unimelb.edu.au}
\affiliation{ARC Centre of Excellence for Dark Matter Particle Physics, School of Physics, The University of Melbourne, Victoria 3010, Australia}

\author{Alexander C.~Ritter}
\email{ritter@unimelb.edu.au}
\affiliation{ARC Centre of Excellence for Dark Matter Particle Physics, School of Physics, The University of Melbourne, Victoria 3010, Australia}

\begin{abstract}
    An ongoing challenge in dark matter direct detection is to improve the sensitivity to light dark matter in the MeV--GeV mass range. One proposal is to dope a liquid noble-element direct detection experiment with a lighter element such as hydrogen. This has the advantage of enabling larger recoil energies compared to scattering on a heavy target, while leveraging existing detector technologies. Direct detection experiments can also extend their reach to lower masses by exploiting the Migdal effect, where a nuclear recoil leads to electronic ionisation or excitation. In this work we combine these ideas to study the sensitivity of a hydrogen-doped LZ experiment (HydroX), and a future large-scale experiment such as XLZD. We find that HydroX could have sensitivity to dark matter masses below 10\,MeV for both spin-independent and spin-dependent scattering, with XLZD extending that reach to lower cross sections. Notably, this technique substantially enhances the sensitivity of direct detection to spin-dependent proton scattering, well beyond the reach of any current experiments.
\end{abstract}

\maketitle


\section{Introduction}

Discerning the origin and nature of dark matter is one of the most important problems in modern particle physics. The hunt for dark matter encompasses a diverse set of search methods; among them, direct detection is a particularly sensitive and flexible method. Decades of progress has resulted in direct detection experiments becoming exquisitely sensitive probes of dark matter parameter space, with large-scale liquid noble-element detectors setting the most stringent constraints on dark matter scattering cross sections for GeV- to TeV-scale masses~\cite{LUX-ZEPLIN:2022qhg,Aprile:2022vux}. While this mass range has attracted great attention -- being the natural home of the weakly-interacting massive particle -- there has recently been much interest in lighter dark matter candidates with mass below a GeV~\cite{Billard:2021uyg}.

Most direct detection experiments lose sensitivity if the dark matter mass is less than a few GeV, since the nuclear recoil energy deposited by the elastic scattering of dark matter on the target nucleus lies below the energy threshold for detection. In recent years, there has been substantial interest in using electromagnetic signatures that can accompany nuclear recoils to extend the reach of experiments to lighter dark matter. Foremost among these is the Migdal effect: the ionisation or excitation of atomic electrons resulting from a nuclear scattering~\cite{Migdal:1939,Migdal:1941,Feinberg:1941,Migdal:1977}.  The significance of the Migdal effect was established in~\cite{Ibe:2017yqa,Dolan:2017xbu} (after earlier initial work~\cite{Vergados:2004bm, Moustakidis:2005gx, Ejiri:2005aj, Bernabei:2007jz}), leading to further theoretical developments and experimental proposals~\cite{Sharma:2017fmo, Essig:2019xkx, Bell:2019egg, Baxter:2019pnz, Knapen:2020aky, GrillidiCortona:2020owp, Liu:2020pat, Liang:2020ryg, Bell:2021zkr,Bell:2021ihi, Knapen:2021bwg, Acevedo:2021kly,Cox:2022ekg,Blanco:2022pkt,Adams:2022zvg,Berghaus:2022pbu}. Although the rate of Migdal events is very small, the Migdal effect can nevertheless provide leading constraints on the light dark matter parameter space~\cite{Dolan:2017xbu, Bell:2019egg, Essig:2019xkx, Bell:2021zkr}, with several experiments conducting dedicated searches~\cite{LUX:2018akb, EDELWEISS:2019vjv, CDEX:2019hzn, XENON:2019zpr, SENSEI:2020dpa, COSINE-100:2021poy, EDELWEISS:2022ktt, DarkSide:2022dhx,SuperCDMS:2023sql}. 

Another, less explored, approach to improve the reach of detectors to light dark matter is via doping with a lighter element, such as hydrogen~\cite{Paschos:1989ei}. This provides a target nucleus with better kinematic matching to light dark matter and, consequently, nuclear recoil energies above the detection threshold. One specific proposal to do this is HydroX~\cite{HydroX:2023}, an upgrade of the LZ experiment that would dope the liquid xenon with molecular hydrogen.

Combining hydrogen-doping with the Migdal effect may ultimately yield the lowest reach in dark matter mass that can be achieved with liquid noble-element detectors (unless dark matter scatters significantly with electrons). In this work, we therefore investigate the sensitivity of hydrogen-doped liquid xenon detectors to low-mass dark matter using the Migdal effect, providing projections for LZ and a next-generation detector such as XLZD. We show that by exploiting the Migdal effect, such detectors can be sensitive to dark matter with mass as low as a few MeV. The improvement in sensitivity to spin-dependent proton scattering is particularly striking, given that a pure xenon target has virtually no sensitivity due to its even number of protons. In this case, the additional reach gained by doping with hydrogen can enable xenon experiments to probe dark matter masses more than an order of magnitude smaller than existing experiments.


\section{The Migdal effect} \label{sec:migdal}

An atomic system perturbed by a nuclear recoil has a small probability of being ionised or excited~\cite{Migdal:1939}. This is known as the Migdal effect and is a consequence of the motion of the recoiling nucleus affecting the electronic wavefunction. The effect has been observed in nuclear decays, where the nuclear recoil is caused by $\alpha$~\cite{PhysRevC.11.1740} and $\beta$~\cite{PhysRevLett.108.243201} emission, and there are efforts currently underway to observe the Migdal effect due to neutron scattering~\cite{Araujo:2022wjh,Nakamura:2020kex}. Such efforts will be useful to validate theoretical calculations of the ionisation probability, especially at high recoil velocities~\cite{Cox:2022ekg}.

The double-differential event rate for a nuclear recoil of energy $E_R$ that gives rise to a Migdal event with electromagnetic energy $E_{EM}$ is given by:
\begin{equation}
    \frac{dR}{dE_{EM}dE_R} = \sum_i \frac{dR}{dE_R} \frac{dP_i}{dE_e} \, , 
    \label{eq:doubleDiff}
\end{equation}
where $\frac{dP_i}{dE_e}$ is the probability of ejecting an electron with energy $E_e$ from the $i^\text{th}$ subshell and $\frac{dR}{dE_R}$ is the differential nuclear recoil rate. The total electromagnetic energy in the event is $E_{EM} = E_e + E_\text{dex}$, where $E_\text{dex}$ is the energy released in the de-excitation of the atomic system. For hydrogen $E_\text{dex}\simeq0$ and $E_{EM} = E_e$. This differs from a Migdal event due to a xenon recoil where $E_\text{dex}$ is the energy of the Auger electrons or photons released in the prompt de-excitation of the xenon ion.

\subsection{Migdal ionisation probability}

The differential Migdal ionisation probability, $dP/dE_e$, is computed from the Migdal transition matrix element:
\begin{equation} \label{eq:Migdal_ME}
    \langle \psi_f | e^{i m_e \bm{v} \cdot \bm{r}} | \psi_i \rangle \,,
\end{equation}
with $\bm{v}$ the nuclear recoil velocity and $|\psi_i\rangle$, $|\psi_f\rangle$ the initial- and final-state electronic wavefunctions, respectively. In the case of atomic hydrogen, where the wavefunctions are known analytically, it is possible to obtain an explicit formula for the ionisation probability. For the small recoil velocities produced by scattering dark matter (more precisely when $v\ll\alpha$, with $\alpha$ the fine-structure constant), the operator in \cref{eq:Migdal_ME} can be approximated by its dipole expansion. In this limit, the ionisation probability is given in terms of the radial wave-functions of the initial-state $1s$ orbital, $R_{1s}$, and the final-state Coulomb wave, $R_{E_e,l'=1}$, by
\begin{align} \label{eq:MigdalAnalytic}
    \frac{dP}{dE_e} = \frac{m_e^2v^2}{3} \, \bigg| \int_0^{\infty} dr\, r^3 R_{E_e,l'=1}(r) R_{1s}(r) \bigg|^2 \,.
\end{align}
This expression is derived in Appendix~\ref{app:MigdalH}.

The above expression for the ionisation probability strictly applies only for atomic hydrogen, not molecular $\mathrm{H}_2$. The Migdal effect in diatomic molecules was investigated in Refs.~\cite{Lovesey:1982,Colognesi:2005,Blanco:2022pkt}, but these works focused on low-energy processes involving excitation to bound excited states. In contrast, we are interested in processes that involve the emission of an ionisation electron and which likely result in the dissociation of the molecule. A precise calculation of this molecular process is highly non-trivial and goes beyond the scope of the present work. We instead adopt the data-driven approach of Ref.~\cite{Liu:2020pat},  which relates the differential Migdal rate in the dipole approximation to the photoabsorption cross section. We use the H$_2$ cross section measurements from Ref.~\cite{Yan_1998}.

Note, however, that the formalism of Ref.~\cite{Liu:2020pat}, which considered atomic systems, does not strictly apply to the molecular case. An additional complication arises in molecular systems where there are two contributions to the Migdal rate: the centre-of-mass recoil (CMR) contribution, which is analogous to the Migdal effect in atoms, and the non-adiabatic coupling (NAC) contribution~\cite{Lovesey:1982,Blanco:2022pkt} arising from corrections to the Born-Oppenheimer approximation for the molecular wavefunctions. Only the CMR contribution can be obtained from the photoabsorption cross section. While the two contributions are comparable in size for excitation to low-energy bound states, the NAC contribution is suppressed at high energies, justifying the use of the data-driven approach for ionisation. This approach is also conservative, as neglecting the NAC contribution may lead to an underestimation of the true Migdal rate for the lowest dark matter masses we consider ($\lesssim$\,10\,MeV). Finally, as an additional cross-check of our results, we also compute the Migdal rate using the atomic expression in \cref{eq:MigdalAnalytic} and find that this is consistent with the data-driven approach to within $\sim30\%$ (see \cref{app:MigdalH}).
%

\subsection{Nuclear recoil rate}

The differential nuclear recoil rate (per unit target mass) is given by
\begin{equation}
    \frac{dR}{dE_R} = \frac{\rho_\chi}{m_\chi m_T} \int_{v_{\mathrm{min}}} d^3v\, \frac{d\sigma}{dE_R} v f(\bm{v})  \,,
\end{equation}
where $\rho_\chi$ is the local density of dark matter, $m_\chi$ is the dark matter mass, and $m_T$ is the target nucleus mass. The local velocity distribution of the dark matter, $f(v)$, is taken to be a truncated Maxwell-Boltzmann distribution, and we adopt the values for the astrophysical parameters given in~\cite{Baxter:2021pqo}. The distribution is integrated above the minimum incoming dark matter velocity that can give rise to a recoil energy of $E_R$ with inelastic energy $\Delta E$: 
\begin{equation}
    v_{\mathrm{min}} = \frac{E_R m_T + \Delta E\mu_{\chi T}}{\sqrt{2 m_T E_R}\mu_{\chi T}} \,,
\end{equation}
where $\mu_{\chi T}$ is the dark matter--target reduced mass. By energy conservation, the inelastic energy is $\Delta E = E_{EM} + E_\mathrm{ion}$, where $E_\mathrm{ion}=$\,13.6\,keV is the ionisation energy of a hydrogen atom. The nuclear recoil energy is constrained to be between:
\begin{equation}
    E_R^\pm = \frac{\mu_{\chi T}^2}{m_T} v_{\mathrm{max}}^2 \left(1-\frac{\Delta E}{\mu_{\chi T} v_{\mathrm{max}}^2} \pm \sqrt{1-\frac{2\Delta E}{\mu_{\chi T} v_{\mathrm{max}}^2}}\right)
\end{equation}
with $v_\mathrm{max}$ the maximum incoming dark matter velocity.

In this work, we consider both spin-independent (SI) and spin-dependent (SD) scattering. The SI and SD differential cross sections for dark matter scattering off a hydrogen atom are
\begin{equation}
    \frac{d\sigma^{\rm SI/SD}}{dE_R}  = \frac{m_p}{2\mu_{\chi p}^2 v^2} \sigma_{\chi p}^{\rm SI/SD} \,,
\end{equation}
with $\sigma_{\chi p}^{\rm SI}$ and $\sigma_{\chi p}^{\rm SD}$ the SI and SD dark matter--proton cross sections, respectively, and $\mu_{\chi p}$ the dark matter--proton reduced mass.


\section{Doping xenon with hydrogen} \label{sec:hydrogen}

While the theoretical benefits of introducing a light dopant into an existing detector are clear, detailed studies of the drift properties, cryogenic properties, and light and charge yields for H$_2$-doped xenon time projection chambers (TPCs) are still underway. In the absence of measurements, we employ simplified treatments of the charge yields, as described below.

In the scattering of sub-GeV mass dark matter, only the ionisation (S2) signal is expected to be detectable, and not the scintillation (S1) signal. We calculate the rate of liberated charge according to
\begin{equation}
    R(N_e) = \int P(N_e\vert \lambda_e(E_R,E_{EM})) \frac{d^2R}{dE_{EM}dE_R} dE_R dE_{EM} \,,
\end{equation}
where $P(N_e\vert \lambda_e(E_R,E_{EM}))$ is the Poisson probability of observing $N_e$ electrons given the expected ionisation $\lambda_e$. This is calculated from the energy deposits via
\begin{equation}
    \lambda_e(E_R,E_{EM}) = E_R Q^{NR}_y(E_R) + E_{EM} Q^{ER}_y(E_{EM}) \,,
\end{equation}
with $Q^{ER/NR}_y(E)$ the charge yield functions.

First, we discuss our treatment of the nuclear recoil charge yield, $Q^{NR}_y(E_R)$. In general, one expects the number of observable quanta to be higher for a recoiling proton than a xenon nucleus, which loses $\sim$80\% of its recoil energy to heat. We follow the approach of HydroX~\cite{HydroX:private} and use a simple form for the nuclear recoil charge yield:
\begin{equation}
    Q_y^{NR}(E_R) = \frac{\alpha}{W_q} \mathcal{L}(E_R) \,,
\end{equation}
where $W_q = 13.7$\,eV is the work function, and $\alpha = 0.67$ gives the fraction of quanta from a recoil that are electrons. The Lindhard factor is taken to be $\mathcal{L}(E_R) = 0.9403{E_R}^{0.01735}$, which comes from SRIM simulations performed by HydroX~\cite{ZIEGLER20101818}. The above value of $\alpha$ corresponds to an NR-like partitioning between light and charge, which should be conservative for an S2-only search. We neglect the recombination of electrons following initial ionisation, since it is expected to give only a small correction to the charge yield for low-energy recoils~\cite{LUX:2015amk, LUX:2016rfb}. Finally, to avoid overestimating the charge yield at low energies, we follow the approach of the Noble Element Simulation Technique (NEST)~\cite{Szydagis:2022} and make the conservative approximation that the charge yield is zero below recoil energies of 0.1\,keV. 

Note that the Migdal signal for dark matter masses below $\sim$100 MeV is dominated by the charge yield from the recoiling Migdal electron; hence, our projected sensitivity at low dark matter masses will not be affected by the above uncertainties associated with the unmeasured charge yield of the recoiling proton. 

For the electron recoil charge yield, $Q_y^{ER}(E_{EM})$, we use the $\beta$~model from NEST v2.3.5~\cite{Szydagis:2022}. This assumes that the charge yield in a liquid xenon TPC is unaffected by its doping with hydrogen, although this was found not to be the case for higher energy recoils in a gaseous xenon TPC~\cite{Tezuka:2004}.

Measurements of light and charge yields in liquid xenon TPCs doped with a light element are in progress, with first observations made of helium recoils in a mini-TPC~\cite{Haselschwardt:2023iqn}. In this work we only consider doping with H$_2$, but helium and deuterium are other potential dopants~\cite{Aalbers:2022dzr}.


\section{Projected Sensitivity} \label{sec:results}

\begin{figure}
    \centering
    \includegraphics[width=0.47\textwidth]{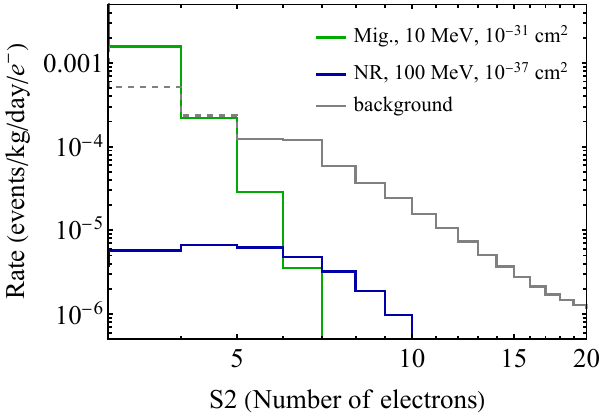}
    \caption{Expected background rate at the LZ experiment (grey), compared with potential signals from dark matter recoiling on hydrogen; the green and blue curves show a Migdal and a nuclear recoil-only signal, respectively. The dashed grey line shows our extrapolation of the LZ background for $\mathrm{S2}<5e^-$.}
    \label{fig:spectra}
\end{figure}

\subsection{S2-only analysis}

A well-established approach to maximise the sensitivity of liquid xenon TPCs to low-energy signals is to perform an analysis that uses only the ionisation (S2) signal. This allows for a lower energy threshold at the expense of discrimination between electron and nuclear recoils. The LZ experiment has published their projected sensitivity to light dark matter using an S2-only analysis in Ref.~\cite{Akerib:2021pfd}. To obtain projections for the improved mass reach that may be achieved through hydrogen doping, we adopt a similar analysis here. 

The LZ analysis in \cite{Akerib:2021pfd} assumed a cut on the S2 signal of $\mathrm{S2}>420$\,phe (equivalent to a threshold of five extracted electrons), and used information on both the magnitude of the S2 signal and its pulse shape to discriminate between signal and background. In our analysis we consider the same $5e^-$ threshold and also investigate the improvement that could be obtained with a $3e^-$ threshold. The latter possibility has also been considered by the HydroX collaboration~\cite{Haselschwardt:2023iqn}, but may come at the expense of reduced trigger efficiency.\footnote{The XENON1T experiment set limits on light dark matter using a $1e^-$ threshold; however, this used data from an R\&D run with a modified trigger and required stringent event selection cuts that severely limited the effective exposure~\cite{XENON:2021qze}.}

We adopt the LZ background model from \cite{Akerib:2021pfd}, and use an S2-binned Poisson likelihood ratio to obtain projected upper limits on the signal cross section. We do not fully incorporate S2 pulse shape information into our analysis, since \cite{Akerib:2021pfd} does not provide the required 2D background distributions. Instead, we assume that the background produced by the anode grid can be removed via pulse shape discrimination, while retaining $\approx100\%$ signal efficiency. A full 2D analysis would be expected to yield improved sensitivity, with pulse shape also providing some discrimination against other backgrounds, in particular from the cathode. 

The background model in \cite{Akerib:2021pfd} is only provided for $\mathrm{S2}\geq5e^-$. For our analysis with a $3e^-$ threshold, we fit the background in the $5e^-$ to $20e^-$ range with a power law and use this to extrapolate to lower S2. We note that single-electron backgrounds can cause a steeply rising background at low thresholds, even after cuts designed to remove them. This was observed by XENON1T below 5 electrons~\cite{XENON:2021qze}, and LUX below 3 electrons~\cite{LUX:2020vbj}. In this work we assume the single-electron background does not contribute significantly above 3 electrons (beyond our naive extrapolation). The S2 distributions of the background and prospective Migdal and nuclear recoil signals from dark matter are shown in \cref{fig:spectra}.

For an S2-only analysis, the absence of an S1 signal means that S1--S2 timing information cannot be used to perform a fiducialization cut in the vertical ($z$) direction.  While some information on $z$ can be gleaned from the S2 pulse width (due to the different drift distances), we follow~\cite{Akerib:2021pfd} and omit this, taking the fiducial mass for S2-only searches in LZ to be 6.2\,t, where the fiducial volume includes the full vertical height of the detector. 

Lastly, we note that injecting kilograms of hydrogen into a detector carries the risk of also introducing tritium, whose decays would add a significant additional source of background. For this background to be subdominant, we assume that the hydrogen used has a tritium concentration of no more than 1 part per $10^{23}$.\footnote{Tritium is created in the atmosphere by cosmic rays, giving rise to typical concentrations of 1 part per $10^{18}$, therefore a significantly depleted source of hydrogen would need to be obtained e.g. from underground gas wells.} 

\subsection{Results}

\begin{figure*}
    \centering
    \includegraphics[width=0.45\textwidth]{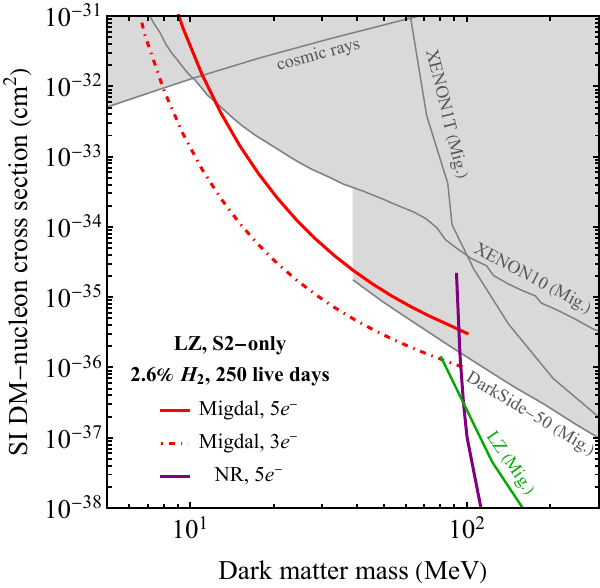}\hfill
    \includegraphics[width=0.45\textwidth]{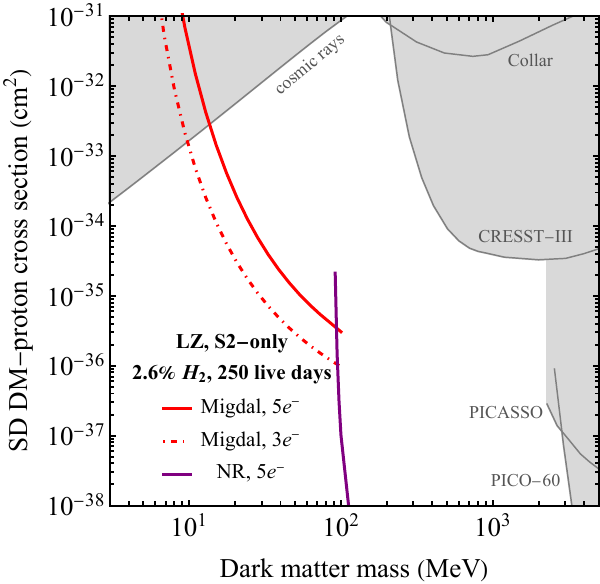}
    \caption{Projected 90\%\,C.L. upper limits on spin-independent nucleon (left) and spin-dependent proton (right) scattering in hydrogen-doped liquid xenon at the LZ experiment. Migdal searches with a $5e^-$ ($3e^-$) threshold are shown in solid (dot-dashed) red, while the purple curve is for a nuclear recoil-only search. The green curve shows the un-doped LZ Migdal projection from~\cite{Akerib:2021pfd}. The grey regions are excluded by previous experiments.}
    \label{fig:H_Mig}
\end{figure*}

We explore the potential of hydrogen doping in both the existing LZ experiment, and also a next-generation xenon experiment, such as XLZD~\cite{Aalbers:2022dzr}, which we denote G3.

For the LZ experiment, we follow the HydroX collaboration~\cite{HydroX:2023} and consider an H$_2$-doped run of 250 live days with a doping fraction of 2.6\% H$_2$ by mole. This corresponds to 2.5\,kg of H$_2$ within the total fiducial mass of 6.2\,t. Data-taking would occur after the conclusion of the planned LZ physics run. 

\Cref{fig:H_Mig} (left) shows the projected sensitivity of LZ to spin-independent nuclear scattering of light dark matter. The green line shows the S2-only Migdal projection from the LZ collaboration in Ref.~\cite{Akerib:2021pfd}, which uses the same background as our analyses but also includes pulse-shape discrimination. We supplement this with new projections assuming hydrogen doping, based on our calculations described above. The red curves show the projected sensitivities of Migdal searches with a $5e^-$ threshold (solid) and a $3e^-$ threshold (dot-dashed). For comparison, the purple line shows our projection for an S2-only nuclear recoil (NR) search with a $5e^-$ threshold (see also the preliminary NR projections from HydroX~\cite{HydroX:2023}). 

The grey region in \cref{fig:H_Mig} has been excluded by previous experiments. There are published limits from Migdal searches by XENON1T~\cite{XENON:2019zpr} and DarkSide-50~\cite{DarkSide:2022dhx}. In addition, data from XENON10~\cite{XENON10:2011prx} has been re-interpreted in~\cite{Essig:2019xkx} to constrain a Migdal signal. The XENON10 data was obtained with a $1e^-$ trigger threshold, resulting in excellent sensitivity at low dark matter masses. Finally, at large cross sections there are limits from cosmic-ray upscattering. These have some model-dependence, and the bound in \cref{fig:H_Mig} assumes fermionic dark matter and a vector mediator that is heavy compared to the momentum exchanged when the DM scatters in the detector~\cite{Dent:2019krz}.

We find that with a $5e^-$ threshold, the Migdal search can improve upon current bounds in a dark matter mass range between 12 and 40\,MeV. The reach of this search improves significantly with a $3e^-$ threshold and could probe new regions of parameter space in the 8\,--\,95\,MeV mass range with cross sections down to $1\times10^{-36} \,\rm{cm}^2$ for $m_\chi=95$\,MeV.

\begin{figure*}
    \centering
    \includegraphics[width=0.45\textwidth]{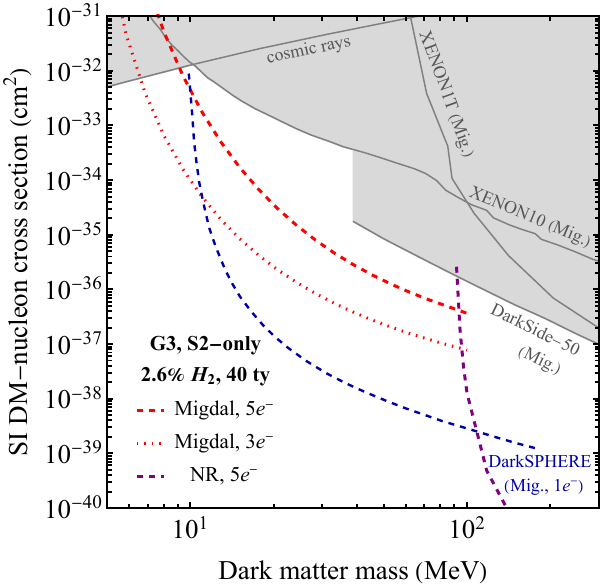}\hfill
    \includegraphics[width=0.45\textwidth]{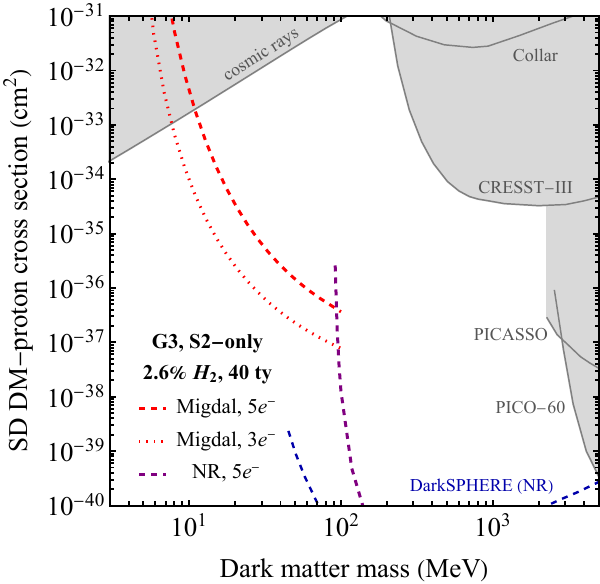}
    \caption{Projected 90\%\,C.L. upper limits on spin-independent nucleon (left) and spin-dependent proton (right) scattering in hydrogen-doped liquid xenon at a future G3 detector (red and purple curves). Also shown are projections for the proposed DarkSPHERE experiment~\cite{NEWS-G:2023qwh} (dashed-blue). The grey regions are excluded by previous experiments.}
    \label{fig:H_Mig_scaled}
\end{figure*}

The prospects are much better for spin-dependent scattering, where hydrogen doping has the potential to significantly improve the sensitivity to low-mass dark matter. Since xenon contains an even number of protons, liquid xenon experiments are primarily sensitive to SD dark-matter--neutron scattering via the naturally occurring isotopes $^{129}$Xe and $^{131}$Xe. While xenon experiments have some limited sensitivity to proton scattering, they are not competitive with experiments containing a fluorine target, such as PICO-60~\cite{PICO:2019vsc}. This situation changes completely with hydrogen doping. The projected sensitivity of hydrogen-doped LZ to SD dark matter--proton scattering is shown in Fig.~\ref{fig:H_Mig_scaled} (right). Again, the red curves correspond to Migdal searches with $5e^-$ and $3e^-$ thresholds, and the purple curve shows an S2-only nuclear recoil search. We see that LZ could explore vast new regions of parameter space, probing dark matter masses more than an order of magnitude lower than the existing limits from CRESST-III~\cite{CRESST:2022dtl}, Collar~\cite{Collar:2018ydf}, PICASSO~\cite{Behnke:2016lsk}, and PICO-60~\cite{PICO:2019vsc}. For large cross sections there is again a model-dependent bound from cosmic ray upscattering; this assumes fermionic dark matter with an axial-vector mediator that is heavy compared to the exchanged momentum~\cite{Dent:2019krz}.

Next, we consider a next-generation (G3) detector. We take such a detector to have a 20\,t fiducial mass of xenon, doped with 2.6\% H$_2$ by mole, corresponding to 7.1\,kg of hydrogen. We assume that the H$_2$ doping would occur at the end of an approximately decade-long primary physics run, with two years of live time while doped. For the G3 detector we use a different background model. In the LZ background model, coherent scattering of $^8$B solar neutrinos already contributed an $\mathcal{O}(1)$ fraction of the background events. Here, we assume that the cathode background can be further mitigated, and calculate the expected background from $^8$B solar neutrinos only. Our sensitivity projections for SI and SD scattering are shown in \cref{fig:H_Mig_scaled}.

Migdal searches at other planned future detectors may also be sensitive to dark matter masses in the 10\,--\,100\,MeV range. Hydrogen doping could also, in principle, be employed in argon TPCs such as DarkSide-20k~\cite{DarkSide-20k:2017zyg}. Another planned experiment with light target nuclei is DarkSPHERE, a 3\,m diameter spherical proportional counter filled with a 90\%-10\% mixture of helium and isobutane ($\mbox{i--C}_{4} \mbox{H}_{10}$), for a total target mass of 27.3\,kg. The dashed blue line in \cref{fig:H_Mig_scaled} (left) shows the DarkSPHERE collaboration's sensitivity projection for a $1e^-$-threshold Migdal analysis~\cite{NEWS-G:2023qwh}. This projection only includes scattering off helium; however, the mass of hydrogen in the detector (as a constituent of the isobutane) will be 2.9\,kg, similar to that in a hydrogen-doped LZ. Accounting for the Migdal signal from scattering off hydrogen in isobutane could extend the DarkSPHERE SI and SD sensitivity to lower masses. Finally, semiconductor detectors may also be sensitive to this parameter space through the Migdal effect~\cite{Essig:2019xkx}, provided that the observed low-energy backgrounds~\cite{Fuss:2022fxe} can be mitigated.


\section{Outlook} \label{sec:outlook}

Hydrogen doping is a promising approach that, when combined with a dedicated search for ionisation signals produced via the Migdal effect, can maximise the reach of xenon TPCs to low-mass dark matter. We have found that employing this technique within the existing LZ detector, as proposed by the HydroX collaboration, could provide sensitivity to dark matter masses below 10\,MeV and explore significant new regions of parameter space. Furthermore, the presence of hydrogen drastically enhances the sensitivity of xenon experiments to spin-dependent proton scattering, where the sub-GeV mass regime has remained largely unexplored to date. Planned future experiments, such as XLZD, could also significantly improve their low-mass reach through hydrogen doping and the Migdal effect.


\begin{acknowledgments}

We thank Hugh Lippincott and the HydroX collaboration for sharing details of their analysis, and Theresa Fruth for helpful discussions regarding the LZ experiment. The work of NFB, MJD, and JLN is supported by the Australian Research Council through the ARC Centre of Excellence for Dark Matter Particle Physics, CE200100008. PC is supported by the Australian Research Council Discovery Early Career Researcher Award DE210100446. ACR is supported by an Australian Government Research Training Program Scholarship.

\end{acknowledgments}


\appendix

\section{Migdal effect in Atomic Hydrogen}
\label{app:MigdalH}

In this appendix we derive the Migdal ionisation probabilities for atomic hydrogen (we work in natural units, $\hbar=c=1$). These were first calculated in the context of neutron scattering in Ref.~\cite{Ruijgrok:1983}. 
The Migdal transition matrix element is
\begin{equation} \label{eq:Migdal_ME_1e}
    M_{n l m}^{E' l' m'} = \big\langle E',l',m' \big| \exp\left(im_e\bm{v}\cdot \bm{r}\right)  \big|n,l,m\big\rangle \,,
\end{equation}
where $\bm{v}$ is the nuclear recoil velocity and $\bm{r}$ is the electron position operator. The initial state electron has principal, angular momentum, and azimuthal quantum numbers $(n,l,m)$; the energy of the final state continuum electron is denoted by $E'$.

As usual, the Migdal operator can be written as a spherical tensor expansion:
\begin{equation}
    \exp(im_e\bm{v}\cdot\bm{r}) = 4\pi \sum_{L,M} i^L j_L(m_evr)Y_L^{M*}(\hat{\bm{v}})Y_L^{M}(\hat{\bm{r}}) \,.
\end{equation}
The matrix element \eqref{eq:Migdal_ME_1e} then becomes
\begin{multline} \label{eq:ME}
    M_{n l m}^{E' l' m'} = \sqrt{4\pi}\sum_{L,M}i^L \sqrt{2L+1} \, Y_L^M(\hat{\bm{v}}) \, d^L_M(l',m';l,m) \\
    \times \int_0^{\infty} dr\, r^2 j_L(m_evr) R_{E',l'}(r) R_{n,l}(r) \,,
\end{multline}
where $j_L$ is the spherical Bessel function and $R_{n,l}(r)$, $R_{E',l'}(r)$ are the radial functions of the initial and final state electrons respectively, with $\langle\bm{r}|n,l,m\rangle=R_{n,l}(r)Y_l^m(\theta,\phi)$. The angular coefficients $d^L_M(l',m';l,m)$ are determined via the Wigner-Eckart theorem,
\begin{multline}
    d^L_M(l',m';l,m) = (-1)^{2l'-m'} \, \sqrt{(2l+1)(2l'+1)} \\
    \times\left(
    \begin{array}{rrr}
        l' & L & l \\
        -m' & M & m
    \end{array}
    \right)
    \left(
    \begin{array}{rrr}
        l' & L & l \\
        0 & 0 & 0
    \end{array}
    \right) \,.
\end{multline}
It is convenient to define the quantisation axis to be aligned with the direction of the recoil velocity, such that
\begin{equation}
    Y_L^M(\hat{\bm{v}}) = \sqrt{\frac{2L+1}{4\pi}} \, \delta_{M,0} \,.
\end{equation}

The relevant initial state is the $1s$ ground state, with radial function
\begin{equation}
    R_{1s}(r) = 2 a_0^{-3/2} e^{-r/a_0} \,,
\end{equation}
and $a_0$ the Bohr radius. The final state continuum wavefunctions are Coulomb waves:
\begin{multline}
    R_{E',l'}(r) = \sqrt{\frac{2}{\pi\alpha a_0}} \frac{1}{(2m_eE')^{1/4}}\frac{1}{r} \,\text{Im} \Big[e^{i \theta(l',\eta,\rho)} \\
    \times (-2i\rho)^{l'+1+i\eta} U(l'+1+i\eta,2l'+2,-2i\rho) \Big] \,,
\end{multline}
with $\rho=r\sqrt{2m_eE'}$, $\eta=-\alpha \sqrt{m_e/(2E')}$, $U(a,b,z)$ the confluent hypergeometric function, and
\begin{equation}
    \theta(l',\eta,\rho) = \rho - \eta\log(2\rho) - \frac{1}{2}l'\pi + \arg(\Gamma(l'+1+i\eta)) \,.
\end{equation}
We normalise the continuum wavefunctions with respect to energy, $\int dr\, r^2 R_{E',l}\,R_{E,l}=\delta(E'-E)$.

In the relevant kinematic regime for non-relativistic dark matter scattering, the recoil velocity satisfies $m_eva_0=v/\alpha\ll1$. In this limit the $L=1$ dipole transition dominates and the differential Migdal ionisation probability simplifies to
\begin{align}
    \frac{dP^\text{(dipole)}}{dE'} = \frac{m_e^2v^2}{3} \, \bigg| \int_0^{\infty} dr\, r^3 R_{E',l'=1}(r) R_{1s}(r) \bigg|^2 \,.
\end{align}

\begin{figure}
    \centering
    \includegraphics[width=0.47\textwidth]{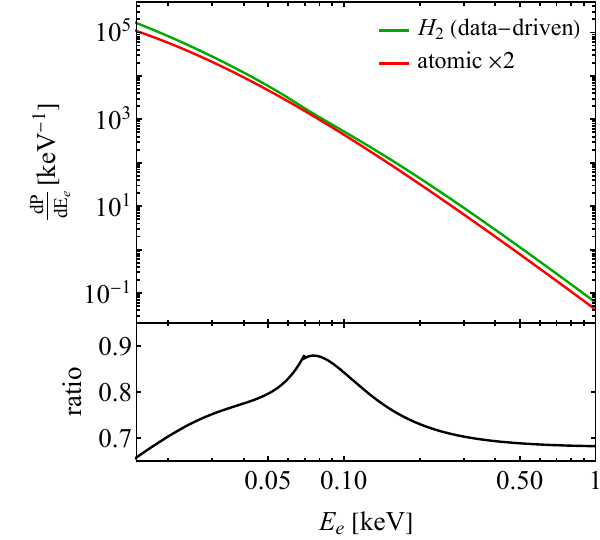}
    \caption{Comparison between the Migdal ionisation probabilities calculated using the data-driven approach (green) and the atomic result in \cref{eq:MigdalAnalytic} (red), both for a nuclear recoil velocity of $v=10^{-3}$. The bottom panel shows the ratio of the atomic result to the data-driven result.}
    \label{fig:mig_prob_comparison}
\end{figure}

In \cref{fig:mig_prob_comparison}, we compare the atomic and molecular hydrogen Migdal ionisation probabilities, with the latter obtained using the data-driven approach of Ref.~\cite{Liu:2020pat}. The nuclear recoil velocity is fixed to $v = 10^{-3}$. We find that they agree to within $\sim$30\%.

It is also straightforward to calculate the integrated Migdal transition probability, including both ionisation and bound excitations,
\begin{align}
     P_\text{integrated} &= 1 - \Big|\big\langle 1s \big| \exp\left(im_e\bm{v}\cdot \bm{r}\right)  \big|1s\big\rangle\Big|^2 \notag \\
     &= 1- \frac{16}{\left(4+(v/\alpha)^2\right)^2} \,.
\end{align}


\bibliography{migdal.bib}

\end{document}